\documentclass[12pt]{article}
\usepackage{amsfonts,amssymb,latexsym}
\newtheorem{theorem}{Theorem}[section]
\newtheorem{lemma}[theorem]{Lemma}
\newtheorem{corollary}[theorem]{Corollary}
\newtheorem{proposition}[theorem]{Proposition}

\newenvironment{proof}
{\par\addvspace{0.3cm}\noindent{\rm Proof. }}
{\nopagebreak\mbox{}\hfill $\Box$\par\addvspace{0.25cm}}

\newcommand{\be}{\begin{equation}}
\newcommand{\ee}{\end{equation}}
\newcommand{\ds}{\displaystyle}

\newcommand{\bq}{\begin{eqnarray}}
\newcommand{\eq}{\end{eqnarray}}
\newcommand{\nn}{\nonumber}
\newcommand{\ba}{\begin{array}}
\newcommand{\ea}{\end{array}}
\newcommand{\tr}{\mathrm{trace}}
\renewcommand{\Re}{\mathrm{Re\,}}
\renewcommand{\Im}{\mathrm{Im\,}}
\newcommand{\wt}{\widetilde}
\newcommand{\wh}{\widehat}
\newcommand{\iv}{^{-1}}
\newcommand{\iy}{\infty}

\newcommand{\noi}{\noindent} 

\newcommand{\twomat}[1]{\left(\ba{cc} #1 \ea\right)}
\newcommand{\twotwo}[4]{\left(\begin{array}{cc}#1&#2\\&\\#3&#4\end{array}\right)}

\newcommand{\C}{{\mathbb C}}
\newcommand{\Z}{{\mathbb Z}}
\newcommand{\T}{{\mathbb T}}

\newcommand{\cS}{\mathcal{S}}
\newcommand{\cR}{\mathcal{R}}
\newcommand{\cQ}{\mathcal{Q}}
\newcommand{\cF}{\mathcal{F}}
\newcommand{\cN}{\mathcal{N}}

\newcommand{\B}{\mathcal{B}}

\newcommand{\eps}{{\varepsilon}}
\renewcommand{\rho}{{\varrho}}
\renewcommand{\kappa}{\varkappa}
\newcommand{\al}{\alpha}
\newcommand{\NN}{^{N\times N}}
\newcommand{\x}[1]{\xi_{#1}}

\begin{document}

\date{}
\title{Asymptotics of block Toeplitz determinants and the classical dimer model}
\author{Estelle L. Basor\thanks{ebasor@calpoly.edu. 
          Supported in part by NSF Grants DMS-0200167 and DMS-0500892.}\\
               Department of Mathematics\\
               California Polytechnic State University\\
               San Luis Obispo, CA 93407, USA
        \and
        Torsten Ehrhardt\thanks{ehrhardt@math.ucsc.edu}\\
     	      Department of Mathematics \\
	      University of California \\
               Santa Cruz, CA-95064, USA}
\maketitle

\begin{abstract}
We compute the asymptotics of a block Toeplitz determinant which arises
in the classical dimer model for the triangular lattice when considering the monomer-monomer
correlation function. The model depends on a parameter interpolating between the square lattice
($t=0$) and the triangular lattice ($t=1$), and we obtain the asymptotics for $0<t\le 1$.
For $0<t<1$ we apply the Szeg\"o Limit Theorem for block Toeplitz determinants.  The main difficulty
is to evaluate the constant term in the asymptotics, which is generally given only in a rather abstract form.
\end{abstract}

 \section{Introduction} This paper is devoted to computing the asymptotic behavior of a certain determinant that arises in the study of the classical dimer model on the triangular lattice. In \cite{FMS},  Fendley, Moessner, and Sondhi study the monomer-monomer correlation function $P^{(mm)}(n)$, which 
 can be expressed by a determinant of a block matrix
  \begin{eqnarray}\label{Pmm}
 P^{(mm)}(n)=\frac{1}{2}\sqrt{\det M_n},\qquad 
M_n= \twotwo{\cR}{\cQ}{\cQ}{\cR}, \label{dimer.matrix}
 \end{eqnarray}
where $\cR$ is an $n\times n$ matrix with entries depending on the difference of their indices,
and $\cQ$ is an $n\times n$ matrix with entries depending on the sum of the indices. 
The entries also depend on a parameter $t$ which interpolates between the square lattice ($t = 0$) and the triangular lattice ($t=1$). The authors computed this correlation function numerically for $t=1$ as the size of the matrix increases and found that it converged to a constant value of around $0.1494\dots$. For all the details of the dimer model computation, the reader is referred to \cite{FMS}.
 
The main purpose of our paper is to compute the determinant asymptotically for all values of $t$ between zero  and one. In fact, we will be able to determine the asymptotics for all complex parameters $t$ with $\Re(t)>0$. This covers all physically interesting cases except $t=0$. 

Our method is to convert the determinant of the block matrix $M_n$ that arises in the dimer problem into a determinant of a block Toeplitz matrix and then to find a way to explicitly compute the asymptotics of the block Toeplitz determinant. 

To be more specific the dimer matrix (\ref{dimer.matrix}) has $n \times n$ matrix blocks $\cal{R}$ and $\cal{Q}$ whose entries are given by 
\bq
 \cR_{jk} &=&  2(-1)^{[(k-j)/2]}R_{k-j+1} +\theta(j-k)t^{j-k-1}\nn \\
 \cQ_{jk} &=& 2i(-1)^{[(j+k)/2]}Q_{n+1-j-k},\nn
\eq
($1\le j,k\le n$) and where the expressions  $R_{k}, Q_{k},$ and $\theta(k)$ are defined by the following.\\
For even $k$ 
\[R_{k} = \frac{1}{8\pi^{2}}\int_{-\pi}^{\pi}\int_{-\pi}^{\pi}\frac{\cos y \cos(kx + y)}{\cos^{2}x  + \cos^{2}y + t^{2}\cos^{2}(x + y)}dydx,\] 
and $Q_{k} = 0.$\\

\noi For odd $k$
\[
Q_{k} =  \frac{1}{8\pi^{2}}\int_{-\pi}^{\pi}\int_{-\pi}^{\pi}\frac{\cos x \cos kx}{\cos^{2}x + \cos^{2}y + t^{2}\cos^{2}(x + y )}dydx,\]
and
\[R_{k}  =  \frac{t}{8\pi^{2}}\int_{-\pi}^{\pi}\int_{-\pi}^{\pi}\frac{\cos(x+y)\cos(kx+y)}{\cos^{2}x + \cos^{2}y + t^{2}\cos^{2}(x + y )}dydx.\]
The expression $\theta(k)$ equals $1$ for $k > 0$ and $0$ otherwise.

In the Appendix we  show that the determinant of the dimer matrix $M_n$ is the same as the determinant of an $n \times n$ block Toeplitz matrix. 
A finite block Toeplitz matrix is one of the form
\[T_{n}(\phi) = (\phi_{j-k}), \quad 0 \leq j,k \leq n -1,\]
where 
\[ \phi_{k} = \frac{1}{2\pi}\int_{0}^{2\pi} \phi(e^{i x})e^{-ik x} \,dx \] 
are the (matrix) Fourier coefficients of an $N \times N$ matrix valued function $\phi$ defined on the unit circle $\T=\{\; z\in\C:\; |z|=1\;\}$. In the dimer case we have $N = 2$, and for $0<t<1$ we obtain
\be
\det M_n=\det T_n(\phi)\quad\mbox{with}\quad
\phi= \twotwo{c}{d}{{\tilde d}}{{\tilde c}}, \label{bl.toe} 
\ee
where 
\bq
c(e^{ix}) &=& \frac{ (t \cos x + \sin^{2}x)}{(t-e^{-ix})\sqrt{t^{2 }+ \sin^{2}x +\sin^{4}x}} 
\\
d(e^{ix}) &=&  \frac{\sin x}{\sqrt{t^{2 }+ \sin^{2}x +\sin^{4}x} }.
\eq
Here and in what follows $\tilde{a}(e^{ix}) = a(e^{-ix}).$

In the case $0<t<1$, we are able to find the asymptotic expansion of the determinant of $T_n(\phi)$ by
using the Strong Szeg\"o Limit Theorem. The main difficulty is that the constant term in the asymptotics for block Toeplitz matrices is generally only given in a rather abstract form. Thus we are required to find a more explicit expression for the constant, which fortunately, in this case, can be done. 
For this purpose we will use an expression for the constant which holds in very special cases and which is due to Widom \cite{W1}. We will also use another determinantal identity which is perhaps new and by the help of which more general (yet special) cases can be reduced to the situation covered by Widom.
We will also give a new proof of Widom's result using the Borodin-Okounkov-Case-Geronimo identity \cite{BO,BS,GC}. All these results are described in the Section \ref{s.detid} and are of independent interest. 

The application to the dimer case still requires a lot of elementary, yet tedious computations which will be done in Sections \ref{s.basic} and \ref{s.det}. The final result is that the limit of the monomer-monomer correlation function
\bq\label{P.mm1}
P^{(mm)}(\iy):=\lim_{n\to\iy} P^{(mm)}(n)
\eq
exists and is given by the formula
\be\label{P.mm}
P^{(mm)}(\iy) =
 \frac{1}{2}\sqrt{\frac{t}{2t(2+t^{2}) + (1+2t^{2})\sqrt{2+t^2}}}.
 \ee
 At this point we have proved this in the case $0<t<1$.
 
In the case $t=1$, we will also obtain this asymptotic formula, which agrees with the numerical calculation \cite{FMS}. In this case we cannot use the Szeg\"o Limit Theorem because the function $c$ has a singularity. This singularity is due to the second term in the definition of the entries of ${\cal R}$ in terms of $R_k$. The entries of $M_n$ are well-defined and analytic for $t\in\C$ with
$\Re(t)>0$. For these values we will actually be able to prove the same asymptotic formula as above.
This will be done in Section \ref{gen.par}. The idea is to transform the matrix $T_n(\phi)$ 
into a matrix which is a finite block Toeplitz matrix plus a certain perturbation (which is of fixed finite rank). A generalization of the block version of the Seg\"o Limit Theorem then gives the expected 
asymptotics. The statement and the proof of this generalized version is given in Section
\ref{s.sz}, and is also of independent interest.


\section{Determinant identities}
\label{s.detid}
 
We begin with some preliminary facts about Toeplitz operators and Toeplitz matrices. Let $\phi \in L^{\infty}(\mathbb{T})^{N \times N}$ be an essentially bounded $N \times N$ matrix valued function defined on the unit circle with Fourier coefficients $\phi_k\in \C^{N\times N}$.
The Toeplitz and Hankel operators are defined on $\ell^{2}(\Z_+)$, $\Z_+=\{0,1,\dots\}$, by means of the semi-infinite infinite matrices
\begin{eqnarray*}
 T(\phi) & =&  (\phi_{j-k}), \,\,\,\,\,\,\,\,\,\,\,\,\,0 \leq j,k < \infty,\\[1ex]
 H(\phi) &=&  (\phi_{j+k+1}), \,\,\,\,\,\,\,\,0 \leq j,k < \infty.
 \end{eqnarray*}
For $\phi, \psi \in L^{\infty}(\mathbb{T})^{N \times N}$ the identities
\bq\label{fT1}
T(\phi \psi) & =& T(\phi)T(\psi) +H(\phi)H({\tilde \psi})\\
H(\phi \psi) & =& T(\phi)H(\psi) +H(\phi)T({\tilde \psi})
\eq
are well-known. It follows from these identities that if  $\psi_{-}$ and $\psi_{+} $ have the property that all their Fourier coefficients vanish for $k > 0$ and $k < 0$, respectively, then 
\bq\label{fT2}
T(\psi_{-} \phi \psi_{+} )& =& T(\psi_{-} )T(\phi) T( \psi_{+} ) ,\\
H(\psi_{-} \phi \tilde{\psi}_{+} )& =& T(\psi_{-} )H(\phi) T( \psi_{+} ) .
\label{THT}
\eq

In the following sections we compute an explicit expression for one of the constants that appears in the version of the Strong Szeg\"o Limit Theorem for matrix-valued symbols, for certain symbols having a special form. In order to state the classic Strong Szeg\"o Limit Theorem for matrix-valued symbols, let $\B$ stand for the set of all function $\phi\in L^1(\T)$ such that the Fourier coefficients satisfy
 \be\label{B}
 \| \phi\|_{\B} :=\sum_{k =-\iy} ^{\iy} |\phi_{k}| + 
 \Big(\sum_{k = -\iy}^{\iy} |k|\cdot|\phi_{k}|^{2}\Big)^{1/2} < \iy.
 \ee
With the norm (\ref{B}) and pointswise defined algebraic operations on $\T$, the set $\B$ becomes a Banach algebra of continuous functions on the unit circle (see, e.g., \cite{BS,W2}).

\begin{theorem}[Szeg\"o-Widom \cite{W2}]
Let $\phi\in \B^{N \times N}$ and assume that the function $\det \phi$ does  not vanish on $\T$ and has winding number zero. Then 
 \[
 \det T_{n}(\phi) \sim G(\phi)^{n}E(\phi) ,\quad \mbox{ as } n\to\iy,
 \]
 where $$G(\phi) = \exp\Big(\frac{1}{2\pi}\int_{0}^{2\pi}\log \det \phi(e^{i x}) dx\Big)$$
 and $$E(\phi) = \det T(\phi)T(\phi^{-1}).$$
 \end{theorem}
 
 In connection with this theorem, let us explain why the definitions of the constants $G(\phi)$ and $E(\phi)$ make sense. First of all, one can show that each 
 nonvanishing function in $\B$ with winding number zero possesses a logarithm which belongs
 to $\B$. Hence $G(\phi)$ is well defined as the geometric mean of $\det \phi$.
 The constant $E(\phi)$ has to be understood as an operator determinant. In fact, we have
 $T(\phi)T(\phi\iv)=I-H(\phi)H(\tilde{\phi}\iv)$ where the product of the Hankel operators is
 a trace class operator. Observe that we have defined $\B$ is such a way that Hankel operators with symbols in $\B$ are Hilbert-Schmidt. For general information about trace class operators and operator determinants\ see, e.g., \cite{GK}.
 
 In the scalar case ($N=1$), there exists an explicit expression for
 $E(\phi)$ given by
 $$
 E(\phi)=\exp\Big(\sum_{k=1}^\iy k [\log \phi]_k[\log\phi]_{-k}\Big).
 $$
 In general ($N\ge2$) a  more explicit expression for $E(\phi)$ does not exist apart from very special cases. For more information about related results we refer the reader to \cite{BS}.

One of the few results concerning an explicit expression for $E(\phi)$ in the matrix case goes back to Widom who showed that if $\phi_{k}$ vanish for $k < n$ or for $k >n$ (for some fixed $n$), then
$E(\phi) = \det T_{n}(\phi^{-1})G(\phi)^{n}$ (see \cite{W1}, Theorem 5.1). Thus the constant $E(\phi)$ is reduced to the computation of a finite determinant, which for small values of $n$  is very computable. If one can somehow reduce the computation of our $E(\phi)$ to this case, then an explicit formula for $E(\phi)$ is possible. The next theorem and its corollary will facilitate such a reduction 
 and since we cannot expect to do it except for very special cases, it requires special assumptions on the symbol.
 
\begin{proposition}
Let $R\in \B^{N\times N}$. Then the operator determinant 
$\det (T(e^R)e^{-T(R)})$ is well defined.
Moreover, 
$f(\lambda)=\det (T(e^{\lambda  R})e^{-T(\lambda R)})$
is an entire function.
\end{proposition}
\begin{proof}
This is proved in \cite[Sect.~7]{E4}.
\end{proof}

In what follows let $I_N$ denote the $N\times N$ identity matrix.

\begin{theorem}\label{d.th}
Let $Q\in \B^{N\times N}$, $\tr \,Q=0$, and $a\in \B$.
Then
\[ \det (T(e^{aI_N + Q})e^{-T(aI_N + Q)}) = \exp\Big( \frac{N}{2}\mathrm{trace\,} H(a)H({\tilde{a})}\Big) 
\cdot \det (T(e^{Q})e^{-T(Q)}).\]
\end{theorem}

\begin{proof}
This proof is modeled on the ones given in \cite{E3,E4} and a more detailed account of why the various determinants and derivatives are defined can be found there. We give a only a sketch here. Let $R=aI_N+Q$ and define
$$f( \lambda) = \det (T(e^{\lambda R})e^{-\lambda T(R)}).$$ 
The function $f(\lambda)$ is analytic on $\C$ and it is nonzero if and only if the Toeplitz operator
$T(e^{\lambda R})$ is invertible. In particular, $f(\lambda)$ is nonzero except on a discrete subset of $\C$. Using the fact that
 $( \log \det F(\lambda))' = \mathrm{trace\,} F'(\lambda)F\iv(\lambda)$ we have that 
 \[
 f'(\lambda)/f(\lambda) =\mathrm{trace\,} \Big(T(Re^{\lambda R})T^{-1}(e^{\lambda R}) - 
 T(e^{\lambda R})T(R) T^{-1}(e^{\lambda R}) \Big).\]
 This implies 
\[ f'(\lambda)/f(\lambda) =\mathrm{trace\,} \Big(T^{-1}(e^{\lambda R})T(Re^{\lambda R}) - 
 T(R) \Big)  .\]
Differentiating again we have that 
\begin{eqnarray*}
(f'(\lambda)/f(\lambda) )'  = \mathrm{trace\,} \Big(T^{-1}(e^{\lambda R}) T(R^{2}e^{\lambda R}) 
- T^{-1}(e^{\lambda R}) T(Re^{\lambda R })T^{-1}(e^{\lambda R}) T(Re^{\lambda R}) \Big).
\end{eqnarray*}
At the points where $f(\lambda)\neq0$, the symbol $\psi^{\lambda} = e^{\lambda R} $ has the property that it factors into a product 
\[ \psi^{\lambda} = \psi_{-}^{\lambda}\psi_{+} ^{\lambda}\]
such that the factors $\psi_{-}$ and $\psi_{+}$ as well as their inverses belong to $\B^{N\times N}$ and
have Fourier coefficients that vanish for $k > 0$ and $k < 0 $ respectively. Thus, by (\ref{fT2}), 
\[T( e^{\lambda R}) = T(\psi_{-}^{\lambda})T(\psi_{+}^{\lambda}). \]
This yields $$T^{-1}( e^{\lambda R}) = T((\psi_{+}^{\lambda})^{-1})T((\psi_{-}^{\lambda} )^{-1}).$$
After simplifying we obtain  
\[(f'(\lambda)/f(\lambda) )' = \mathrm{trace\,} \Big(T((\psi_{-}^{\lambda} )^{-1}R^{2}\psi_{-}^{\lambda}) -T^{2}((\psi_{-}^{\lambda} )^{-1}R\psi_{-}^{\lambda})\Big)
= \tr(H(\rho)H({\tilde{\rho}})) \]
where 
$$\rho = (\psi_{-}^{\lambda} )^{-1}R\psi_{-}^{\lambda} = aI_N + 
(\psi_{-}^{\lambda} )^{-1}Q\psi_{-}^{\lambda}.$$
Now define 
$$g( \lambda) = \det (T(e^{ \lambda Q})e^{-\lambda T( Q)})$$ and compute the above expression once again with $R$ replaced by $Q$ (i.e., put $a=0$). Then the difference 
$$(f'(\lambda )/f(\lambda))' - (g'(\lambda)/g(\lambda))' = N\cdot\mathrm{trace\,} H(a)H(\tilde{a}).$$
The reason for this is that if we expand $$\mathrm{trace\,} H(\rho)H({\tilde{\rho}})$$ we  have four terms.
A term of the form $H(aI_N)H(\widetilde{(\psi_{-}^{\lambda} )^{-1}Q\psi_{-}^{\lambda}})$ has trace zero since each matrix Fourier coefficient of $(\psi_{-}^{\lambda} )^{-1}Q\psi_{-}^{\lambda}$ has this property and multiplication by a matrix coefficient of $aI_N$ corresponds to a scalar multiplication.
The term $$\tr\, H((\psi_{-}^{\lambda} )^{-1}Q\psi_{-}^{\lambda})
H(\widetilde{(\psi_{-}^{\lambda} )^{-1}Q\psi_{-}^{\lambda}})$$ 
cancels with the corresponding term for $g$ since the corresponding functions 
$\psi_{-}^{\lambda}$ (obtained from  $e^{\lambda a I_N+\lambda Q}$ and $e^{\lambda Q}$, resp., by factorization) only differ by a scalar function. 
Finally the ``$N$'' comes from the fact that $aI_N$ is a 
$N \times N$ matrix.

Thus 
$$\log f(\lambda) - \log g(\lambda)$$ has a constant second derivative and satisfies 
$$(\log f - \log g)(0) = 0 \,\,\,\, \mbox{and} \,\,\,\,\,(f'/f)(0) -( g'/g)(0) = 0.$$ 
From this the theorem follows easily.
\end{proof}

\begin{corollary}\label{c2.2}
Let $Q\in \B^{N\times N}$, $\tr\, Q=0$, and $a\in \B$. 
Then
$$
E(e^{aI_N + Q}) = \exp\Big(N\cdot \mathrm{trace\,} H(a)H({\tilde{a})}\Big)\cdot E(e^{Q}).
$$
\end{corollary}
\begin{proof}
We can write
\begin{eqnarray*}
E(e^{aI_N + Q})&=&
\det T(e^{aI_N + Q})T(e^{-aI_N-Q})\\
&=&
\det T(e^{aI_N + Q})e^{-T(aI_N+Q)} \cdot \det e^{T(aI_N + Q)}T(e^{-aI_N-Q})
\end{eqnarray*}
and apply the previous theorem twice.
\end{proof}

The following theorem and proposition yields an alternative proof to Widom's result in the case where either the positive or negative Fourier coefficients vanish for all but a finite number of indices. We give a  proof that is based on the Borodin-Okounkov-Case-Geronimo (BOCG) identity \cite{BO,GC}. Widom's result is stated in our Theorem \ref{t.w}.

\begin{proposition}\label{p2.3}
Suppose that $\psi\in\B^{N\times N}$ such that both $T(\psi)$ and $T(\tilde{\psi})$ are invertible on $(\ell^2(\Z_+))^N$.
Then
\bq\label{f2}
\det T_n(\psi\iv) &=& \frac{E(\psi)}{G(\psi)^n}\cdot 
\det\left(I-H(z^{-n}\psi)T\iv(\tilde{\psi})H(\tilde{\psi} z^{-n})T\iv(\psi)\right).
\eq
\end{proposition}
\begin{proof}
The invertibility of $T(\psi)$ and $T(\tilde{\psi})$ implies the invertibility of $\psi$ and  the existence of a left and a right  canonical factorization (in $\B^{N \times N}$)
$$
\psi\iv=u_{-}u_{+}=v_{+}v_{-}.
$$
Proceeding as in the proof of the BOCG-identity one obtains
$$
\det T_{n}(\psi\iv) =G(\psi\iv)^{n}\cdot\det P_{n}T(v_{+}\iv) T(u_{-})T(u_{+})T(v_{-}\iv)P_{n},
$$
where $P_n$ is the projection $\{x_k\}_{k=0}^\iy\mapsto \{x_0,\dots,x_{n-1},0,0, \dots\}$ acting on 
$\ell^2(\Z_+)$. Using the identity 
\begin{eqnarray}
\det P_{n}A P_{n} &=& \det (Q_{n}+P_{n}A P_{n})=\det (Q_{n}+A P_{n})\nn\\
&=& \det A\cdot \det (A\iv Q_{n}+P_{n})=
\det A\cdot\det (Q_{n}A\iv Q_{n})\label{f.K}
\end{eqnarray}
in which $Q_{n}=I-P_{n}$ and $A$ is an invertible operator of the form identity plus trace class, it follows that
\begin{eqnarray*}
\lefteqn{\det P_{n}T(v_{+}\iv) T(u_{-})T(u_{+})T(v_{-}\iv)P_{n}}
\hspace{6ex}\\[1ex]
&=&
\det T(v_{+}\iv) T(u_{-})T(u_{+})T(v_{-}\iv)\cdot
\det Q_{n}T(v_{-})T(u_{+}\iv)T(u_{-}\iv)T(v_{+})Q_{n}.
\end{eqnarray*}
The first term on the right is easily seen to be $\det T(\psi)T(\psi\iv)$ while the 
second term is equal to
$$
\det T(z^{-n} v_{-}u_{+}\iv)T(u_{-}\iv v_{+}z^{n}) =
\det (I-H(z^{-n} v_{-}u_{+}\iv)H(\tilde{u}_{-}\iv \tilde{v}_{+}z^{-n}))
$$
where we use $z^{-n} v_{-}u_{+}\iv u_{-}\iv v_{+}z^{n}=I_{N} $.
Now we substitute $u_{+}\iv=\psi u_{-}$ and $\tilde{u}_{-}\iv=\tilde{u}_{+}\tilde{\psi}$
and pull out the remaining factors from the Hankel operators (see (\ref{THT})), which give the inverses of the
Toeplitz operators.
\end{proof}

\begin{proposition}\label{p2.4}
Let $\psi\in \B^{N\times N}$ be invertible. Then $\det T_n(\psi\iv)=0$ if and only if the operator
$$
\twomat{T(\wt{\psi}) & H(\wt{\psi}z^{-n})\\ H(\psi z^{-n}) & T(\psi)}
$$ 
has a non-trivial kernel (or, equivalently, a non-trivial cokernel).
\end{proposition}
\begin{proof}
The proof relies on the general fact that the kernel (resp.~cokernel) of an operator
$PAP+Q$ is trivial if and only if the kernel (resp.~cokernel) of the operator
$QA\iv Q+P$ is trivial, where $A$ is assumed to be an invertible operator, $P$ is a projection  operator and $Q=I-P$. Indeed, this follows from the fact that both operators are equivalent, i.e., there exist
invertible $S_1$ and $S_2$ such that $PAP+Q=S_1(QA\iv Q+P) S_2$. (One can also say that
$PAP$ and $QA\iv Q$ are equivalent after extension. See (\ref{f.K}) for the underlying line of argumentation.)

We apply this statement in the setting $A=M(\psi\iv)$ being the Laurent operator acting  on
$(\ell^2(\Z))^N$, and $P$ being the projection $\{x_k\}_{k\in\Z}\mapsto \{x_k\}_{k=0}^{n-1}$ on $(\ell^2(\Z))^N$.
\end{proof}

\begin{theorem}[Widom \cite{W1}]\label{t.w}
Let $\psi\in \B^{N\times N}$ be such that the function $\det \psi$ does not vanish on $\T$ and has winding number zero. Assume that $\psi_k=0$ for all $k> n$ or that $\psi_{-k}=0$ for all $k> n$. Then
\bq
E(\psi)=G(\psi)^n \det T_n(\psi\iv).
\eq
\end{theorem}
Note the above result holds also for $n=0$ when stipulating $\det T_0(\psi\iv)=1$.

\begin{proof}
The winding number condition implies that both $T(\psi)$ and $T(\psi\iv)$
are Fredholm operators with index zero. Since one can show that $T(\psi\iv)$ and $T(\wt{\psi})$ are equivalent after extension,  $T(\wt{\psi})$ is also a Fredholm operator with index zero.

Hence if $E(\psi)\neq0$, then both $T(\psi)$ and $T(\psi\iv)$ (and also $T(\wt{\psi})$) are invertible.
Now Proposition \ref{p2.3} settles the assertion.

If $E(\psi)=0$ then (since $T(\psi)T(\psi\iv)=I -$compact), the product
$T(\psi) T(\psi\iv)$ has a non-trivial kernel and cokernel. This implies that 
$T(\psi)$ has non-trivial kernel and cokernel, or, that 
$T(\psi\iv)$ (hence $T(\wt{\psi})$) has non-trivial kernel and cokernel.

Consider the case that $\psi_{k}=0$ for all $k>n$. (The other case can be treated analogously.)
Then the operator considered in Proposition \ref{p2.4} takes the form
$$
\twomat{ T(\wt{\psi}) & H(\wt{\psi} t^{-n}) \\ 0 & T(\psi)}.
$$ 
In the case where $T(\wt{\psi})$ has a non-trivial kernel, the operator above has a non-trivial
kernel. However, if $T(\psi)$ has a non-trivial kernel (hence a non-trivial cokernel),
the above operator has a non-trivial cokernel, too.
\end{proof}


\section{A generalized Szeg\"o-Widom limit theorem}
\label{s.sz}

When dealing with the case of parameters $t\notin (0,1)$ in Section \ref{gen.par}, we are going to employ a generalization of the matrix version of the Szeg\"o-Widom limit theorem. The reader who is not interested in the details of this case can skip this section.

In the analysis of this case we are led to compute the asymptotics of the determinants of block Toeplitz matrices which are perturbed in some way by trace class operators. Although this generalization is very much straightforward, we give the details of the proof because our matrices depend analytically on the parameter $t$ and we want to show that the limiting constant also depends analytically on $t$. This will enable us to identify the constant.

The kind of sequences for which we are going to establish the limit theorem are the $N\times N$ block
versions of the sequences described as follows. Recalling first the definition of the Banach algebra $\B$ of smooth functions given in the previous section, let $\cF$ stand for the set of all
sequences $(A_n)_{n=1}^\iy$ of $n\times n$ matrices which are of the from
$$
A_n=T_n(a)+P_nKP_n+W_n L W_n+C_n
$$
where $a\in \B$, the operators $K$ and $L$ are trace class operators on $\ell^2$,  and $C_n$ are $n\times n$ matrices tending to zero in the trace norm. The set of such sequences $(C_n)_{n=1}^\iy$
will be denoted by $\cN$. The operators  $P_n$ and $W_n$ acting on $\ell^2$ are defined by
\bq
P_n&:& \{x_k\}_{k=0}^\iy \mapsto \{x_0,x_1\dots,x_{n-1},0,0,\dots\},\nn\\
W_n&:& \{x_k\}_{k=0}^\iy \mapsto \{x_{n-1},x_{n-2},\dots,x_{0},0,0,\dots\},\nn
\eq
and $P_nKP_n$ and $W_n L W_n$ are identified with $n\times n$ matrices in the natural way.

Now we are precisely in the setting considered in \cite{E4}. (There is a change of notation, namely, 
that our $\B$ corresponds to $\cS$ of \cite{E4}.) The set $\cF$ is
a Banach algebra with algebraic operations defined elementwise and a norm 
$$
\|(A_n)\|_{\cF}= \|a\|_{\B}+\|K\|_1+\|L\|_1+\sup_{n\ge 1}\|C_n\|_1
$$
where $\|\cdot\|_1$ refers to the trace norm. The subset $\cN$ is a closed two-sided ideal of
$\cF$. 

Let $\mathcal{G}\B^{N\times N}$ stand for the group of all invertible elements in the Banach algebra $\B^{N\times N}$, and denote by $\mathcal{G}_1\B^{N\times N}$ the connected component of 
$\mathcal{G}\B^{N\times N}$ containing the identity element. We remark that for
$a\in \mathcal{G}_1\B\NN$ the constant
$$
G(a)=\exp\Big(\frac{1}{2\pi} \int_0^{2\pi} \log \det a(e^{ix})\, dx\Big)
$$
is well-defined \cite[Sec.~6]{E4}. 

In what follows we are going to consider elements $(A_{n,t})_{n=1}^\iy\in\cF\NN$ which depend analytically on a parameter $t\in \Omega$.
By this we mean that the map $t\mapsto (A_{n,t})_{n=1}^\iy\in\cF\NN$ is an analytic $\cF\NN$-valued
function. If
\be\label{f.15}
(A_{n,t})_{n=1}^\iy = (T_n(a_t) + P_n K_t P_n+W_n L_t W_n +C_{n,t})_{n=1}^\iy \in \mathcal{F}^{N\times N}
\ee
this is equivalent to requiring that the maps $t\mapsto a_t$, $t\mapsto K_t$, $t\mapsto L_t$,
and $t\mapsto (C_{n,t})$ are analytic (because of the definition of the norm in $\cF\NN$).

\begin{theorem}\label{t3.1}
Let $\Omega$ be an open subset of $\C$. For each $t\in \Omega$
let $(A_{n,t})_{n=1}^\iy\in \mathcal{F}^{N\times N}$, and assume that the map
$t\in\Omega\mapsto (A_{n,t})_{n=1}^\iy \in \cF\NN$ is analytic. Moreover, suppose that
$a_t\in \mathcal{G}_1\B\NN$ where $a_t$ is given by (\ref{f.15}).
Then for each $t\in \Omega$ the limit
\bq
E_t= \lim_{n\to \iy} \frac{\det A_{n,t}}{G(a_t)^n}
\eq
exists, the convergence is locally uniform on $\Omega$, and 
$E_t$ depends analytically on $t$.
\end{theorem}
\begin{proof}
By the Vitali-Porter Theorem on induced convergence \cite[Chap.~9]{Bu}, it suffices to show that 
 the sequence $ G(a_t)^{-n}\cdot\det A_{n,t}$ converges pointwise and is locally uniformly bounded on $\Omega$.

Because of the assumption $a_t\in\mathcal{G}_1\B^{N\times N}$, we have a pointwise representation of the form
$$
a_t =e^{b_1}\cdots e^{b_R}, \qquad b_1,\dots,  b_R\in  \B^{N\times N}
$$
with $R$ possibly depending on $t$. 
A moment's thought reveals that for each point $t_0\in \Omega$ one can find a representation of the above kind  
on a sufficiently small neighborhood $\Omega(t_0)$ of $t_0$ such that $b_1,\dots ,b_R$ depend analytically on $t$.
In  what follows we will consider any fixed, but arbitrary point $t_0\in\Omega$ and a corresponding $\Omega(t_0)$.

For $t\in \Omega(t_0)$ consider
$$
B_{n,t}= e^{-T_n(b_1)}\cdots e^{-T_n(b_R)} \Big( T_n(a_t)+P_nK_tP_n +W_nL_tW_n+ C_{n,t}\Big).
$$
The determinant of $B_{n,t}$ equals $ G(a_t)^{-n}\cdot\det A_{n,t}$ since
the trace of  $T_n(b_1)+\dots +T_n(b_R)$ equals $n$ times the zero-th Fourier coefficient of 
$\tr(b_1+\dots +b_R)=\log \det a_t$. Moreover, since $(B_{n,t})$ is built from elements in $\mathcal{F}^{N\times N}$ depending
analytically on $t$, it also belongs to $\cF\NN$ and depends analytically on $t$.

Proceeding as in \cite[Prop.~9.2]{E4}, one can show that the sequence $B_{n,t}$ is of the form
$$
B_{n,t}= P_n+P_n\widehat{K}_tP_n+W_n\widehat{L}_tW_n+\widehat{C}_{n,t}
$$
with $\widehat{K}_t,\widehat{L}_t$ being trace class, $(\widehat{C}_{n,t})_{n=1}^\iy$ belonging to $\mathcal{N}^{N\times N}$, and with analytic dependence.
Applying the Lemma 9.3 of \cite{E4}, it follows that 
$$
\lim_{n\to\iy }\det B_{n,t}= \det (I+\wh{K}_t)\cdot\det (I+\wh{L}_t) 
$$
pointwise. Since $|\det(I+A)|\le \exp\|A\|_1$ for trace class operators $A$, it follows easily that 
$\det B_{n,t}$ is locally uniformly bounded.
\end{proof}

One might think of giving another simpler proof of the previous theorem by considering
the limit of the quotient
$$
\frac{\det A_{n,t}}{\det T_n(a_t)} = 
\det( P_n +T_n\iv(a_t) P_n KP_n + W_n T_n\iv(\tilde{a}_t)  L W_n).
$$
In order to make this work one needs the assumption that $T_n(a_t)$ is stable (i.e. uniformly invertible
as $n\to\iy$). It is known that this equivalent to the operators $T(a_t)$ and $T(\tilde{a}_t)$ both being
invertible. Unfortunately, the assumption on $a_t$ made in the theorem only guarantees that 
these operators are Fredholm with index zero.


\section{The basic computations for $0<t<1$}
\label{s.basic}

The first step in the calculation of the asymptotics of the correlation  function in the dimer model is  to show how the  block matrix considered by the authors in \cite{FMS} has a determinant that can be computed using the theorems of the previous section. As mentioned in the introduction, in \cite{FMS} the authors show the correlation is one-half the square root of the determinant of the matrix $M_n$ given in (\ref{dimer.matrix}).

In the Appendix A we show how to first convert the above coefficients as Fourier coefficients of certain functions in the case $0<t<1$. From this we are able to identify the symbol of the block Toeplitz matrix with $N =2$,  and 
obtain that 
\be\label{f.16}
\det M_n= \det T_n(\phi)  \quad \mbox{ with } \quad 
\phi= \twotwo{c}{d}{{\tilde d}}{{\tilde c}} \label{bl.toe-c} 
\ee
where 
\bq\label{f.17}
c(e^{ix}) &=& \frac{ t \cos x + \sin^{2}x}{(t-e^{-ix})\sqrt{t^{2 }+ \sin^{2}x +\sin^{4}x}}, 
\\ \label{f.18}
d(e^{ix}) &=&  \frac{\sin x}{\sqrt{t^{2 }+ \sin^{2}x +\sin^{4}x} }.
\eq
Recall that $\tilde{a}(e^{i\theta}) = a(e^{-i\theta}).$
Since we now have a block Toeplitz matrix our next step is to apply the Szeg\"o Limit Theorem.
For $0 < t < 1$ the symbol $\phi$ is a smooth function. If we compute the determinant of the symbol we find that  after simplifying, 
\bq\label{dphi}
\det \phi = c {\tilde c} - d {\tilde d} =  
\frac{1}{  t^{2} -2t\cos x + 1} 
\eq
and thus $\det \phi$ does not vanish and has winding number zero.  In addition the geometric mean
$G(\phi)$ is one since
\[ \frac{1}{2\pi}\int_{-\pi}^{\pi} \log \det \phi(e^{ix})\,dx 
= -\frac{1}{2\pi}\int_{-\pi}^{\pi} \log ((1 - te^{ix})(1 - te^{-ix})) dx  = 0.\]
Hence the Szeg\"o-Widom Limit Theorem implies that the asymptotics are given by 
\be \det T_{n}( \phi) \sim E(\phi),\qquad n\to\iy, \ee
and the limit of the monomer-monomer correlation function (\ref{P.mm1}) exists and equals
\bq\label{P.mm.E}
P^{(mm)}(\iy) =\frac{1}{2}\sqrt{E(\phi)}.
\eq

Note this agrees with the result from \cite{FMS} that the determinants should approach a constant.
Now we wish to apply Corollary \ref{c2.2} and Theorem \ref{t.w} to identify the constant $E(\phi)$.
We write 
\[\phi = \sigma   \twotwo{p}{q}{{\tilde q}}{{\tilde p}}\]
where 
\be\label{def:pq}
p(e^{ix}) = ( t \cos x + \sin^{2} x)(t - e^{ix}), \qquad q(e^{ix})  = \sin x ( 1 -2t \cos x + t^{2}), 
\ee
and 
\bq\label{def:sigma}
\sigma(e^{ix}) &=& ( t^{2 } + \sin^{2}x  + \sin^{4}x)^{-1/2}(1 -2t\cos x + t^{2})^{-1}.
\eq
The following lemma is needed in order to verify the assumptions of Corollary \ref{c2.2}.

\begin{lemma}\label{l3.1}
For $0<t<1$, the matrix function $\phi$ can be written as $\phi=e^{aI_2 + bQ}$
with $a,b\in \B$, $Q\in \B^{2\times  2}$, $\tr \, Q=0$. In particular,
$$
a = -\frac{1}{2}\log (1 - 2t \cos x +t^{2}) +\pi i, \qquad
Q =  \twotwo{\frac{p-{\tilde p}}{2}}{q}{{\tilde q}}{\frac{{\tilde p}-p}{2}}.
$$
\end{lemma}
\begin{proof}
Clearly, $a$ and $Q$ satisfy the stated conditions. Let us mention that 
$b$ is given by
$$
b = \frac{1}{\Delta}\left(-a+\pi i+\log\sigma+\log\left(-\frac{p+\tilde{p}}{2}-\Delta \right)\right),
\qquad
\Delta=\sqrt{\frac{(p-\tilde{p})^{2}}{4}+q\tilde{q}},
$$
and let us first show that $b\in\B$ (if the logarithm is chosen appropriately). 
Notice that
$$
\frac{p+\tilde{p}}{2}=( t \cos x + \sin^{2} x)(t - \cos x),\quad
\Delta=i \sin x \sqrt{ (1- 2t\cos x+t^{2})^2+(t\cos x+\sin^{2} x)^2}.
$$
Thus for $\alpha(x)=-\frac{p+\tilde{p}}{2}-\Delta$, we have
$\Im\alpha(x)<0$ for $0<x<\pi$, $\Im\alpha(x)>0$ for $-\pi<x<0$,
and both $\alpha(0)=t(1-t)$ and $\alpha(\pi)=t(1+t)$ are real positive.
Hence $\alpha(x)$ is a continuous function with winding number zero,
which possesses a continuous logarithm which we normalize in such a way that 
the logarithm is real positive for $x=0$ and $x=\pi$. With this normalization
the numerator in the expression for $b$ becomes zero at $x=0$ and $x=\pi$,
which cancels with the zero in the denominator $\Delta$ (as one can easily show). Hence $b\in\B$.

From the definition of $b$ we further conclude that 
$$
e^{\Delta b}=e^{-a}\sigma\left(\frac{p+\tilde{p}}{2}+\Delta\right).
$$
We claim that
$$
e^{-\Delta b}=e^{-a}\sigma\left(\frac{p+\tilde{p}}{2}-\Delta\right).
$$
This can be seen most easily by considering the product and noting that
$$
\frac{(p+\tilde{p})^{2}}{4}-\Delta^{2}=
( t^{2 } + \sin^{2}x  + \sin^{4}x)(1-2t\cos x+t^{2})
=e^{2a}\sigma^{-2}.
$$
It follows that
$$
\cosh(\Delta b)=e^{-a}\sigma\,\frac{p+\tilde{p}}{2},\qquad
\sinh(\Delta b)=e^{-a}\sigma\Delta.
$$
It can be verified straightforwardly that
\[ \phi =  \sigma \left(\frac{p+{\tilde p}}{2} I_2 + Q\right),
\qquad
Q^{2} = -(\det Q)I_2=\Delta^{2 }I_2.
\]
Hence
$$
e^{bQ}=\cosh(b\Delta)I_2+\frac{\sinh(b\Delta)}{\Delta} Q.
$$
Combining all this yields 
$$
e^{bQ}=e^{-a}\sigma\left(\frac{p+\tilde{p}}{2}I_2+Q\right)
$$
and thus $\phi=e^{{aI_2+bQ}}$.
\end{proof}

In the following  computation the factorization of the function $t^2 +\sin^2 x+\sin^4 x$,
which appears in the function $\sigma$, will play an important role. We make a substituion
$z=e^{ix}$ and can rewrite
$$
t^2+\sin^2 x+\sin^4 x = \frac{1}{16}(z^4-8z^2+(14+16 t^2) -8 z^{-2}+z^{-4}).
$$
Making the obvious substitution $z=y^2$ we can now factor
\bq
\lefteqn{y^{2}-8y+(14+16t^{2})-8y\iv+y^{-2}}\hspace*{15ex}\nn\\
&=&\xi_{1}\iv\xi_{2}\iv(1-\xi_{1}y)(1-\xi_{2}y)(1-\xi_{1} y\iv)
(1-\xi_{2}y\iv),\label{poly}
\eq
where $\xi_{1}$ and $\xi_{2}$ are defined by
\be
\label{f.xi}
\xi_{1}:=2+ \mu -  2\,\sqrt{1 - t^2 + \mu},
\qquad
\xi_{2}:=2- \mu -  2\,\sqrt{1 - t^2 - \mu}
\ee
with 
\be\label{f.mu}
\mu:=\sqrt{1 - 4t^2}.
\ee Notice that $|\xi_{1}|<1$ and $|\xi_{2}|<1$. Their inverses are given by
\be
\label{f.xi-iv}
\xi_{1}\iv=2+ \mu +2\,\sqrt{1 - t^2 + \mu},
\qquad
\xi_{2}\iv=2- \mu +  2\,\sqrt{1 - t^2 - \mu}\,\,.
\ee
If $1/2<t<1$ then $\xi_1$ and $\xi_2$ are complex conjugate of each other, whereas
if $0<t<1/2$ they are distinct real numbers. Also notice that 
\bq\label{f.xi+iv}
\x1+\x1\iv=4+2\mu,\qquad
\x2+\x2\iv=4-2\mu.
\eq
Indeed, in order to verify formula (\ref{poly}) it suffices to remark 
that the right hand side equals
$$
(\x1+\x1\iv-y-y\iv)(\x2+\x2\iv-y-y\iv)=16-4\mu^2-8(y+y\iv)+(y+y\iv)^2
$$
which is the same as the left hand side.

\begin{proposition}\label{p3.2}
For $0<t<1$, we have
\bq
E(\phi) &=&
(1-\xi_{1}^{2})(1-\xi_{2}^{2})(1-\xi_{1}\xi_{2})^{2}(1-t^{2}\xi_{1})(1-t^{2}\xi_{2})E(\sigma\iv \phi),\\[1ex]
G(\sigma\iv \phi) &=& \frac{1}{16\xi_{1}\xi_{2}}.
\eq
\end{proposition}
\begin{proof}
First of all notice that, by Lemma \ref{l3.1}, $\phi=e^{a_{1}I_2+bQ}$ and
$\sigma\iv \phi=e^{a_{2}I_2+bQ}$
with
$$
a_{1}=-\frac{1}{2}\log(1-2t\cos x+t^{2})+\pi i,\qquad
a_{2}=a_{1}-\log\sigma.
$$
We apply Corollary \ref{c2.2} twice (with $bQ$ instead of $Q$) and obtain
$$
E(\phi)=\exp\Big(2\,\tr(H(a_{1})H(\tilde{a}_{1})-H(a_{2})H(\tilde{a}_{2}))\Big) E(\sigma\iv \phi).
$$
Introducing $\alpha_{1}=\log(1-2 t\cos x+t^{2})$ and $\alpha_{2}=\log(t^{2}+\sin^{2}x+\sin^{4}x)$,
we have $a_{1}=-\alpha_{1}/2+\pi i$, $a_{2}=a_1+\alpha_1+\alpha_2/2 = (\alpha_{1}+\alpha_{2})/2+\pi i$ 
(see (\ref{def:sigma})), and thus
$$
\exp\Big(2\,\tr(H(a_{1})H(\tilde{a}_{1})-H(a_{2})H(\tilde{a}_{2}))\Big) \hspace{3in}
$$
$$=\exp\Big(-\frac{1}{2}\,\tr(H(\al_{1})H(\tilde{\al}_{2})+
H(\al_{2})H(\tilde{\al}_{1})+H(\al_{2})H(\tilde{\al}_{2}))\Big).
$$
With the substitution $z=e^{{ix}}$ we can write
$$
\al_{1}(z)=\log((1-tz)(1-tz\iv)),\qquad
\al_{2}(z)=\log((z^{4}-8z^{2}+(14+16t^{2})-8z^{-2}+z^{-4})/16).
$$
Since we can decompose
$$
z^{4}-8z^{2}+(14+16t^{2})-8z^{-2}+z^{-4} =\xi_{1}\iv\xi_{2}\iv(1-\xi_{1}z^{2})(1-\xi_{2}z^{2})(1-\xi_{1} z^{-2})
(1-\xi_{2}z^{-2}),
$$
we have the Fourier coefficients ($k\neq 0$)
$$
[\al_{1}]_{k}=-\frac{t^{|k|}}{|k|},\quad
[\al_{2}]_{2k}=-\frac{\xi_{1}^{|k|}+\xi_{2}^{|k|}}{|k|},\quad
[\al_{2}]_{2k+1}=0.
$$
We obtain
$$
\tr H(\al_{1})H(\tilde{\al}_{2})=\tr H(\al_{2})H(\tilde{\al}_{1})=
\sum_{k=1}^{\iy}
t^{2k}\,\frac{\xi_{1}^{k}+\xi_{2}^{k}}{k}=
-\log((1-t^{2}\xi_{1})(1-t^{2}\xi_{2}))
$$
and
$$
\tr H(\al_{2})H(\tilde{\al}_{2})=
\sum_{k=1}^{\iy}
\frac{2(\xi_{1}^{k}+\xi_{2}^{k})^{2}}{k}=
-2\log((1-\xi_{1}^{2})(1-\xi_{2}^{2})(1-\xi_{1}\xi_{2})^{2}).
$$
This implies
$$
E(\phi)=
(1-\xi_{1}^{2})(1-\xi_{2}^{2})(1-\xi_{1}\xi_{2})^{2}
(1-t^{2}\xi_{1})(1-t^{2}\xi_{2})E(\sigma\iv \phi).
$$

In order to compute $G(\sigma\iv \phi)$ notice that 
$$
\det (\sigma\iv \phi) = \frac{\det \phi}{\sigma^{2}}=
(t^{2}+\sin^{2}x+\sin^{4}x) (1-2t\cos x+t^{2}) =\exp(\al_{2}+\al_{1}).
$$
The arithmetic mean of $\al_{1}$ is zero while the arithmetic mean of
$\al_{2}$ is $\log(1/(16\x1\x2))$ as can be seen from the factorization.
\end{proof}

Clearly, the previous proposition reduces the computation of $E(\phi)$ to the computation of
$E(\psi)$ where
\be\label{f.xxx-1}
\psi:= \sigma^{-1}\phi= \twotwo{p}{q}{{\tilde q}}{{\tilde p}}.
\ee
Since the Fourier coefficients $\psi_k$ vanish if $|k| > 3$,
the constant  $E(\psi)$ can be computed from Theorem \ref{t.w} with $n=3$:
\bq\label{f.xxx}
E(\psi)=G(\psi)^3\det T_3(\psi^{-1}) \,\,.
\eq
We know in addition $G(\psi)=G(\sigma\iv \phi)$ so all that remains is the computation of the determinant of  a block $3 \times 3$ Toeplitz matrix whose symbol is the inverse of $\psi^{-1}$.
The next section is devoted to this task because even the determinant of a $3 \times 3$ block matrix (which is in fact $ 6 \times 6$ in size) can be difficult to compute.


\section{The computation of $\det T_3(\psi\iv)$}
\label{s.det}

The symbol $\psi^{-1}$ is given by
\bq\label{def:ab}
\psi^{-1} =   \eta \twotwo{{\tilde p}}{{\tilde q}}{ q}{ p}
=: \twotwo{{\tilde a}}{{ \tilde{b}}}{ b}{ a}
\eq
where $\eta$ is the even function
\[ \eta = (p\tilde{p}-q\tilde{q})\iv= (\det (\sigma\iv \phi))\iv = (1 -2t \cos x + t^{2})\iv(t^{2} + \sin^{2}x +\sin^{4}x)\iv.\]
Using the fact that $b$ is an odd function, it follows that the block Toeplitz matrix $T_3(\psi\iv)$  has the following structure:
\[ T_3(\psi\iv)=\left( \begin{array}{cc|cc|cc}
a_0& 0& a_1&b_1&a_2& b_2 \\[1ex]
0 & a_0 & -b_1 & a_{-1}& -b_2 & a_{-2} \\[1ex]
\hline\rule{0ex}{3ex}
a_{-1} & - b_1 & a_0 & 0 & a_1 & b_1 \\[1ex]
b_1&  a_1 &0& a_0 & - b_1 & a_{-1} \\[1ex] 
\hline\rule{0ex}{3ex}
a_{-2} &-b_2 & a_{-1} & - b_1 & a_0 &0 \\[1ex]
b_2 & a_2 & b_1 & a_1 & 0 & a_0
\end{array}
\right) \] 
What we will next attempt is to show is that the special form of this matrix allows us to reduce its determinant to that of a $ 2 \times 2 $ scalar matrix. The first thing to notice is that if $c_{1} = \det A$ and $c_{2} = \det B$ then the above matrix has the form
\[ \left( \begin{array}{ccc}
a_{0}I_2 & A & B\\
c_{1}A^{-1} & a_{0}I_2 & A\\
c_{2}B^{-1} & c_{1}A^{-1} & a_{0}I_2
\end{array} \right)
\]
Now if we factor out the $a_{0}$ terms and multiply the top block row from the left by $c_{1}a_{0}^{-1}A^{-1}$ and substract this from the second block row we are left with
\[ \left( \begin{array}{ccc}
I_2 & a_{0}^{-1}A & a_{0}^{-1}B\\
 0& (1- c_{1}a_{0}^{-2})I_2 & a_0\iv A - c_{1}a_{0}^{-2}A^{-1}B\\
c_{2}a_{0}^{-1}B^{-1} & c_{1}a_{0}^{-1}A^{-1} & I_2
\end{array} \right).
\]
Except for the factor $a_{0}^{6}$ this has the same determinant as the one of interest.
Multiplying the top block row on the left by
$c_{2}a_{0}^{-1}B^{-1}$ and then subtracting yields in the same manner a determinant equal to that of the matrix
\[ \left( \begin{array}{ccc}
I_2 & a_{0}^{-1}A & a_{0}^{-1}B\\
 0& (1- c_{1}a_{0}^{-2})I_2 & a_0\iv A - c_{1}a_{0}^{-2}A^{-1}B\\
0& c_{1}a_{0}^{-1}A^{-1} -c_{2}a_{0}^{-2}B^{-1}A & (1- c_{2}a_{0}^{-2})I_2
\end{array} \right).
\]
The determinant of the above is given by
\[ \det ((1- c_{1}a_{0}^{-2})(1- c_{2}a_{0}^{-2})I_2-( c_{1}a_{0}^{-1}A^{-1} -c_{2}a_{0}^{-2}B^{-1}A)(a_0^{-1}A - c_{1}a_{0}^{-2}A^{-1}B))
\]
or
\[ 
\det ((1 - 2c_{1}a_{0}^{-2}-c_{2}a_{0}^{-2})I_2 + c_{2}a_{0}^{-3}B^{-1}A^{2}+ c_{1}^{2}a_{0}^{-3}A^{-2}B).
 \]
Now any $2 \times 2$ matrix of the form $(\det M) M^{-1} + M$ is a constant times the identity. This is easy to see if one thinks of the inverse of the matrix. In fact if the matrix is given by
\[ M=\twotwo{m_1}{m_{12}}{m_{21}}{m_{2}} \] then the constant is $m_1+m_2$. The matrix
\[
c_{2}B^{-1}A^{2}+ c_{1}^{2}A^{-2}B 
\]
is of this form and it follows that the matrix given in the last determinant expression is a constant times the identity.
Recalling the definition of the terms $A$ and $B$ we see that this constant is 
$$
a_{0}^{-3}( a_{0}^{3} - 2c_{1}a_{0} -c_{2}a_{0} + a_{-2}(a_{1}^{2} - b_{1}^{2}) + 2b_{1}b_{2}(a_{1}+a_{-1} )+a_{2}(a_{-1}^{2} -b_{1}^{2})).
$$
Since we originally had  a factor of $a_{0}^{6}$ the $a_{0}^{-3}$ term cancels (since this comes from a $2 \times 2$ matrix). Thus, at this point,  we have shown that 
\bq\label{f.Lam}
\det T_3(\psi\iv) &=& \Lambda^2
\eq
 with 
\bq \label{constant}
\Lambda &:=&
a_{0}^{3} - 2c_{1}a_{0} -c_{2}a_{0} + a_{-2}(a_{1}^{2} - b_{1}^{2}) + 2b_{1}b_{2}(a_{1}+a_{-1} )+a_{2}(a_{-1}^{2} -b_{1}^{2}).
\eq

Now we will investigate the Fourier coefficients.
Going back to (\ref{def:pq}) and (\ref{def:ab}), the functions
$a$ and $b$ are given by 
\bq
 b(e^{ix}) = \frac{\sin x}{t^{2} + \sin^{2}x +\sin^{4}x},\qquad
 a(e^{ix}) = \frac{(t \cos x + \sin^{2}x)}{(t - e^{-ix})(t^{2} + \sin^{2}x + \sin^{4}x)}.
 \eq
 With the substitution $z = e^{ix}$ the function $b$ can be written as
 \bq
 b(z) &=& \frac{8i(z^{-1}-z)}{z^{4}-8z^{2}+(14+16t^{2})-8z^{-2}+z^{-4} }\nn\\
& =&\frac{8i\x1\x2(z^{-1}-z)}{(1-\xi_{1}z^{2})(1-\xi_{2}z^{2})(1-\xi_{1} z^{-2})
(1-\xi_{2}z^{-2})}.\nn
\eq

The Fourier coefficients of this can be easily computed and a little algebra shows that $b_{2} =0$ and
\[ b_{1} = \frac{-8i\xi_{1}\xi_{2}}{(1 + \xi_{1})(1+\xi_{2})(1 - \xi_{1}\xi_{2})} .\]
With $b_{2} = 0$ we also can simplify the constant (\ref{constant}) to
\be \label{constant2}
\Lambda=a_{0}^{3} - 2(a_{1}a_{-1} +b_{1}^{2})a_{0} -a_{2}a_{-2}a_{0} + a_{-2}(a_{1}^{2} - b_{1}^{2}) + a_{2}(a_{-1}^{2} -b_{1}^{2}).
\ee
The function $a$ can be rewritten as
\bq
 a(z) &=& \frac{8 t (z+z\iv)-4(z+z\iv)^2}{(t-z\iv)(z^{4}-8z^{2}+(14+16t^{2})-8z^{-2}+z^{-4})}.\nn
 \eq
Using the factorization employed for finding the coefficients for $b$ it can be shown in a messy, yet elementary computation that the following holds:
Let 
\be\label{f.al}
\al := \frac{4\xi_{1}\xi_{2}}{(1 - \xi_{1}\xi_{2})(\xi_{1} - \xi_{2})}
\ee
and 
\bq\label{f.k-x}
 k(x) &:=& \frac{ x^{2} -4x -1}{(1 - t^{2}x)(1 - x^{2})} 
 =\frac{1}{1-t^2 x}\left(\frac{2}{1+x}-\frac{2}{1-x}-1\right),
\\
\label{f.l-x}
l(x) &:=& \frac{(1 - 2t^{2})x^{2}+ (-2-2t^{2})x +1}{(1 - t^{2}x)(1 -x^{2})} =k(x)+\frac{2}{1-x}.
\eq
Then  the Fourier coefficients of $a$ are given by
\begin{eqnarray*}
 a_{0} &=& t\al(\xi_{1}k(\xi_{1})- \xi_{2}k(\xi_{2 })),\\
 a_{1} &=& \al( l(\xi_{1} )- l(\xi_{2}) ),\\
 a_{-1}& =& \al (\xi_{1}l(\xi_{1}) - \xi_{2}l(\xi_{2})),\\
 a_{2} &=& t\al(k(\xi_{1})- k(\xi_{2})),\\
 a_{-2} &=& t\al(\xi_{1}^{2}k(\xi_{1})- \xi_{2}^{2}k(\xi_{2 })) .
  \end{eqnarray*}
With this notation our constant (\ref{constant2}) becomes
\[
\begin{array}{c}
\Lambda = \al^{3}(\xi_{1}-\xi_{2})^{2}\Big( t^{3}\xi_{1}k(\xi_{1})^{2}k(\xi_{2}) -t^{3}\xi_{2}k(\xi_{2})^{2}k(\xi_{1})
\\
+tk(\xi_{1})l(\xi_{2})^{2}-
tk(\xi_{2})l(\xi_{1})^{2} +4(1+\xi_{1})^{-2}(1+\xi_{2})^{-2}(tk(\xi_{1})(\xi_{1}+1)^{2} -tk(\xi_{2})(\xi_{2}+1)^{2})\Big).
\end{array}
\]
Simplifying this expression is the last remaining task.


For simplicity we will assume that $1/2<t<1.$ In this case the roots $\xi$ and $\x2$ are conjugate and $\mu$ is complex. The case $0 < t < 1/2$ is similar, but we omit the details. The answer however is valid for all $0<t<1$ because of analyticity of the expressions with respect to $t$.
We begin by proving that the expression
\be \label{constant3}
\begin{array}{c}
t^{3}\xi_{1}k(\xi_{1})^{2}k(\xi_{2}) -t^{3}\xi_{2}k(\xi_{2})^{2}k(\xi_{1})
 +tk(\xi_{1})l(\xi_{2})^{2}-tk(\xi_{2})l(\xi_{1})^{2}
\\[2ex] \displaystyle
+\frac{4tk(\xi_{1})}{(1+\xi_{2})^{2}} -\frac{4tk(\xi_{2})}{(1+\xi_{1})^{2}} 
\end{array}
\ee
is  equal to 
\be
 \frac{8\mu}{(1+ \xi_{1})(1+\xi_{2})\sqrt{\omega}}
\ee
where 
\be
 \omega := (1 - t^{2}\xi_{1})(1-t^{2}\xi_{2}) >0
\ee

First recall the identities
\be\label{alg1a}
\x1+\x1\iv= 4+2\mu,\qquad
\x2+\x2\iv= 4-2\mu,\qquad
\mu^2=1-4t^2
\ee
from which it is easy to conclude that 
\be\label{alg1ab}
(\x1-1)(\x1\iv-1)(\x2-1)(\x2\iv-1) = 16 t^2.
\ee
In order to evaluate the expression in (\ref{constant3}) we first simplify
\be\label{alg1}
t^{3}\xi_{1}k(\xi_{1})^{2}k(\xi_{2}) -t^{3}\xi_{2}k(\xi_{2})^{2}k(\xi_{1})
 +tk(\xi_{1})l(\xi_{2})^{2}-tk(\xi_{2})l(\xi_{1})^{2}.
\ee
Using (\ref{f.k-x}) and (\ref{f.l-x}) the expression
$$t^{3}\xi_{1}k(\xi_{1})^{2} -tl(\xi_{1})^{2}$$
becomes
\begin{eqnarray*}
t^{3}\xi_{1}k(\xi_{1})^{2} -t\left(k(\xi_{1}) +\frac{2}{1-\xi_{1}}\right)^{2}
&=&
t^{3}\xi_{1}k(\xi_{1})^{2} -tk(\xi_{1})^{2} -\frac{4tk(\xi_{1})}{1-\xi_{1}}-\frac{4t}{(1-\xi_{1})^{2}}.
\end{eqnarray*}
A little more algebra shows that this is the same as
$$t k(\xi_{1})\left(-\frac{2}{1+\xi_{1}}-\frac{2}{1-\xi_{1}} + 1\right)
-\frac{4t}{(1-\xi_{1})^{2}}.$$
Since we have only used the identity for $l(x)$ in terms of $k(x)$ the above is also true if we replace 
$\xi_{1}$ with $\xi_{2}$.  Using this last above equation (for both $\xi_{1}$ and $\xi_{2}$ terms) we have that (\ref{alg1}) is given by
\bq
t k(\xi_{1})k(\xi_{2})\left(-\frac{2}{1+\xi_{1}}-\frac{2}{1-\xi_{1}} +\frac{2}{1+\xi_{2}}+\frac{2}{1-\xi_{2}}  \right) -
\frac{4t k(\xi_{2})}{(1-\xi_{1})^{2}} + \frac{4t k(\xi_{1})}{(1-\xi_{2})^{2}}\nn
\eq
which is of course
\be\label{f25}
4t k(\xi_{1})k(\xi_{2})\left(\frac{\xi_{2}^2- \xi_{1}^2} {(1 - \xi_{1}^{2})(1 - \xi_{2}^{2})}\right) -
\frac{4t k(\xi_{2})}{(1-\xi_{1})^{2}} + \frac{4t k(\xi_{1})}{(1-\xi_{2})^{2}}.
\ee
We now proceed to simplify the above.
From (\ref{alg1a}) it is easy to show that 
\[
(1-\mu \xi_{1})^{2} = 4\xi_{1}(1 - t^{2}\xi_{1}),\qquad
(1+\mu \xi_{2})^{2} = 4\xi_{2}(1 - t^{2}\xi_{2}),
\]
which  using (\ref{alg1ab}) implies
\bq\label{f.om2}
\omega=\frac{(1-\mu\x1)^2(1+\mu\x2)^2}{16\x1\x2}=
\frac{(1-\mu\x1)^2(1+\mu\x2)^2(\x1-1)^2(\x2-1)^2}{(16\x1\x2 t)^2}.
\eq
Moreover,  using $2(1-\mu\x1)=1-\x1^2+4\x1$ and $2(1+\mu\x2)=1-\x2^2+4\x2$ 
(which also follows from (\ref{alg1a}))
we conclude
$$
k(\xi_{1}) = \frac{-8\xi_{1}}{(1-\mu \xi_{1})(1-\xi_{1}^{2})},\qquad
k(\xi_{2}) = \frac{-8\xi_{2}}{(1+\mu \xi_{2})(1-\xi_{2}^{2})}.
$$
With this we have that
\[
k(\xi_{1})k(\xi_{2})  = \frac{64\xi_{1}\xi_{2}}{(1-\mu \xi_{1})(1+\mu \xi_{2})(1-\xi_{1}^{2})(1-\xi_{2}^{2})}.
\]
Using (\ref{f.om2}) it then follows that
\[
k(\xi_{1})k(\xi_{2})  = \frac{4}{t\sqrt{\omega}(1+ \xi_{1})(1+ \xi_{2})}.
\]
Hence the term (\ref{alg1}) (which is also (\ref{f25})) is
\[
 \frac{16(\xi_{2}^{2}-\xi_{1}^{2})}{\sqrt{\omega}(1+ \xi_{1})(1+ \xi_{2})(1-\xi_{1}^{2})(1-\xi_{2}^{2})}
 -\frac{4t k(\xi_{2})}{(1-\xi_{1})^{2}} + \frac{4t k(\xi_{1})}{(1-\xi_{2})^{2}} ,
\]
 and the expression for the constant (\ref{constant3})
 is the above plus the additional terms of 
 \[\frac{4t k(\xi_{1})}{(1+\xi_{2})^{2}}  -\frac{4t k(\xi_{2})}{(1+\xi_{1})^{2}}.
\]
In other words we still need to evaluate 
\be\label{alg4}
\begin{array}{c}\displaystyle
\frac{16(\xi_{2}^{2}-\xi_{1}^{2})}{\sqrt{\omega}(1+ \xi_{1})(1+ \xi_{2})(1-\xi_{1}^{2})(1-\xi_{2}^{2})}
\\[3ex]
\displaystyle
 -\frac{4t k(\xi_{2})}{(1-\xi_{1})^{2}} + \frac{4t k(\xi_{1})}{(1-\xi_{2})^{2}} +\frac{4tk(\xi_{1})}{(1+\xi_{2})^{2}}  -\frac{4tk(\xi_{2})}{(1+\xi_{1})^{2}}.
 \end{array}
\ee
If we combine the terms 
\[
\frac{4t k(\xi_{1})}{(1-\xi_{2})^{2}} +\frac{4tk(\xi_{1})}{(1+\xi_{2})^{2}}
\]
we have
\[
\frac{8tk(\xi_{1})(1+\xi_{2}^{2})}{(1-\xi_{2})^{2}(1+\xi_{2})^{2}}
\]
which becomes after using the identity for $k(\xi_{1})$
\[
\frac{-64 t\x1(1+\xi_{2}^{2})}{(1-\mu\x1)(1-\xi_{1}^{2})(1-\xi_{2})^{2}(1+\xi_{2})^{2}}.
\]
Using (\ref{f.om2}) and (\ref{alg1a}) this becomes
\[
\frac{-4(\x2\iv+\xi_{2})(1+\mu\x2)}{\sqrt{\omega}(1+\xi_{1})(1-\xi_{2})(1+\xi_{2})^{2}}
=
\frac{-8(2 - \mu)(1+\mu \xi_{2})}{\sqrt{\omega }(1+\xi_{1})(1+\xi_{2})(1-\xi_{2}^{2})}
\]
or
\[
\frac{-16+4\mu(2-4\x2+4\mu\x2)}{\sqrt{\omega }(1+\xi_{1})(1+\xi_{2})(1-\xi_{2}^{2})}
\]
and finally
\[
\frac{-16}{\sqrt{\omega} (1+\xi_{1})(1+\xi_{2})(1-\xi_{2}^{2})} +\frac{4\mu}{\sqrt{\omega} (1+\xi_{1})(1+\xi_{2})}.
\]
We still have another term same as the above but with the $\xi_{1}$ and $\xi_{2}$ switched, $\mu$ replaced by $-\mu$ and an additional sign change.
However when we add this new term to the above and then substitute in (\ref{alg4}) we have that our constant (\ref{constant3}) is 
\be \label{final}
\begin{array}{c}\displaystyle
\frac{16(\xi_{2}^{2}-\xi_{1}^{2})}{\sqrt{\omega}(1+ \xi_{1})(1+ \xi_{2})(1-\xi_{1}^{2})(1-\xi_{2}^{2})}
\\[3ex] 
\displaystyle
+\frac{-16}{\sqrt{\omega} (1+\xi_{1})(1+\xi_{2})(1-\xi_{2}^{2})} +\frac{16}{\sqrt{\omega} (1+\xi_{1})(1+\xi_{2})(1-\xi_{1}^{2})}
\\
\displaystyle
 + \frac{8\mu}{\sqrt{\omega} (1+\xi_{1})(1+\xi_{2})} 
 \end{array}
\ee
which nicely reduces to 
\[
\frac{8\mu}{\sqrt{\omega} (1+\xi_{1})(1+\xi_{2})} 
\]
as promised.
Thus the constant (\ref{constant2})  is 
\be\label{f.48x}
\Lambda=
\frac{8\mu \alpha^{3} (\xi_{1}-\xi_{2})^{2}}{\sqrt{\omega} (1+\xi_{1})(1+\xi_{2})} 
\quad\mbox{with}\quad\al = \frac{4\xi_{1}\xi_{2}}{(1 - \xi_{1}\xi_{2})(\xi_{1} - \xi_{2})}.
\ee

Now we can put the final pieces together.
\begin{theorem}
For $0<t<1$, we have 
\be E(\phi) = \frac{t}{2t(2+t^{2})+(1+2t^{2}) \sqrt{2+t^{2}}}.
\ee
\end{theorem}
\begin{proof}
We have from Proposition 3.2 and formulas (\ref{f.xxx-1}), (\ref{f.xxx}), (\ref{f.Lam}), and
(\ref{f.48x}) that 
\[
E(\phi) = (1-\xi_{1}^{2})(1-\xi_{2}^{2})(1-\xi_{1}\xi_{2})^{2}
(1-t^{2}\xi_{1})(1-t^{2}\xi_{2}) (16\xi_{1}\xi_{2})^{-3}
\]
\[ 
\times \left(\frac{8\mu \alpha^{3} (\xi_{1}-\xi_{2})^{2}}{\sqrt{\omega} (1+\xi_{1})(1+\xi_{2})}\right)^{2}.
\]
This can be reduced to 
\[ \frac{4^{5}t^{2}\mu^{2}\xi_{1}^{4}\xi_{2}^{4}}{(1-\xi_{1}^{2})(1-\xi_{2}^{2})(1-\xi_{1}\xi_{2})^{4}(\x1-\x2)^{2}}.\]
This uses only elementary algebra and formula (\ref{alg1ab}).

To evaluate the remaining terms we note (\ref{alg1a}) implies that 
\bq \label{alg 1c}
\frac{(1-\x1\x2)(\x1-\x2)}{\x1\x2} &=& -\x1+\x2-\x1\iv+\x2\iv\;=\; -4\mu.
\eq
and
\[ (1+\xi_{1})(1+\xi_{1}^{-1})(1+\xi_{2})(1+\xi_{2}^{-1}) = (6-2\mu)(2+2\mu)=16(2+t^2) .
\] 
{}From the last equation together with (\ref{alg1}) we can conclude that 
\[(1-\x1^{2})(1-\x2^{2}) = 16\sqrt{t^{2}(2+t^{2}) }\x1\x2.\]
Thus from the line above and (\ref{alg 1c}) it follows that 
\[ E(\phi) =\frac{4t\xi_{1}\xi_{2}}{\sqrt{2+t^{2}}(1-\xi_{1}\xi_{2})^{2}}.\]
To complete the computation notice again (\ref{alg1a}), hence
\[
(\x1+\x1\iv)(\x2+\x2\iv)= (4+2\mu)(4-2\mu) = 12+16t^{2}
\]
Also we know
\[(\x1\iv-\x1)(\x2\iv-\x2) = 16\sqrt{t^{2}(2+t^{2}) },\]
so we can check that
\[\x1\x2 + 1/\x1\x2 = 6 + 8t^{2} + 8\sqrt{t^{2}(2+t^{2})}.\]
This yields that
\[\frac{\xi_{1}\xi_{2}}{(1-\xi_{1}\xi_{2})^{2}} = 
\frac{1}{ 4 + 8t^{2} + 8\sqrt{t^{2}(2+t^{2})}}.\]
From this the theorem follows.
\end{proof}


\section{The asymptotics for more general parameters}
\label{gen.par}

In the previous sections we have shown that 
\bq\label{f.48}
 \det M_n \sim E(\phi), \qquad n\to\iy,
\eq
if $0<t<1$ and we have identified the constant $E(\phi)$.
The goal of this section is to show that (\ref{f.48}) holds for all values of parameters $t$
belonging to the set 
$$
\Omega:=\Big\{\;  t\in\C    \;:\: \Re(t)>0 \; \Big\}
$$
under the assumption that the entries of $M_n$ are continued by analyticity in $t$ onto $\Omega$.
Indeed, if we recall the definition of the entries of $M_n$ in terms of $R_k$ and $Q_k$ (as stated in the
introduction) 
one only needs to observe that the denominator of the fraction under the integrals is nonzero
as long as $t^2$ is not real non-positive number, i.e., if $\Re(t)\neq 0$.

Since we have related $M_n$ to $T_n(\phi)$ by simple row and column operations (see (\ref{bl.toe}) or
(\ref{f.16})) we are going to elaborate on $T_n(\phi)$, the entries of which also depend analytically on $t$. Recall that $\phi$ is given by
$$
\phi=\twotwo{c_t}{d_t}{\tilde{d_t}}{\tilde{c}_t}
$$
with 
\bq
c_t(e^{ix}) & = &  \ds\frac{ t \cos x + \sin^{2}x}{(e^{-ix}-t)\sqrt{t^{2 }+ \sin^{2}x +\sin^{4}x}},\nn\\
d_t(e^{ix}) &=& \ds\frac{\sin x}{\sqrt{t^{2 }+ \sin^{2}x +\sin^{4}x} }.\nn
\eq
Moreover, we introduce the function
\bq
e_t^{+}(e^{ix}) & = & \ds \frac{ t \cos x + \sin^{2}x}{(e^{-ix}-t)\sqrt{t^{2 }+ \sin^{2}x +\sin^{4}x}} 
-\frac{1}{e^{-ix}-t}.\nn
\eq

\begin{proposition}
For each $t\in\Omega$ the functions $d_t$ and $e_t^+$ belong to $\B$, and the dependence on $t$ is analytic.
\end{proposition}
\begin{proof}
The statement concerning the function $t$ is east to see. Indeed, the function $\sin^2 x+\sin^4 x$ belongs to $\B$ and has its spectrum equal to $[0,2]$. By functional calculus we can define the
reciprocal of the square-root of $t^2+\sin^2x+\sin^4x$ whenever $t\in \C\setminus
[-i\sqrt{2},i\sqrt{t}]$ and also at $t=\iy$. This domain (extended with the point at infinity)
is simply connected and we can choose an analytic branch. Cleary, we choose the branch in such a way that the square-root is positive for real, positive $t$. 

In regard to the function $e_t$ the assertion is obvious except for $|t|=1$.
We rewrite this  function as
\bq
e_t(e^{ix})&=&\frac{(t\cos x+\sin^2 x) -\sqrt{t^2+\sin^2 x+\sin^4 x}}{(e^{-ix}-t)\sqrt{t^2+\sin^2 x+\sin^4 x}}.\nn
\eq
We are going to show that term $e^{-ix}-t$ cancels with a term in the numerator for $|t|=1$,
$\Re(t)>0$, and that the resulting function belongs to $\B$ and depends analytically on $t$.
To see this  rewrite the numerator (with the substitution $z=e^{ix}$) as a quarter times
$$
n(t,z)=2t(z+z\iv)-(z-z\iv)^2-\sqrt{z^4-8z^2+14+16t^2-8z^{-2}+z^{4}}.
$$
Allow, for a moment, $z$ to take complex values in a neighborhood of the unit circle. More specifically, 
it is easy to see that 
for each $\eps>0$ there exists a $\delta>0$ such that the function under the square-root is 
nonzero (and of course analytic in both $z$ and $t$) whenever $\Re(t)>\eps$, $|t|< 1+\eps$,
$1-\delta<|z|<1+\delta$. Denote the set of all $(t,z)$ satisfying these conditions by $U_\eps$.
Hence $n(t,z)$ is analytic on $U_\eps$. Consider the subset $M_\eps=\{(t,z)\in U_\eps\;:\; z=1/t\}$. 
Then
$$
n(t,t\iv )=(t^2-t^{-2}+4)-\sqrt{(t^2-t^{-2}+4)^2}.
$$
For $t=1$, keeping in mind the proper choice of the branch of the square-root, this equals zero.
Hence by analytic continuation and since $M_\eps$ is connected and the function under the square-root 
does not vanish on $M_\eps$ it follows that $n(t,z)=0$ for all $(t,z)\in M_\eps $. Hence
$n(t,z)=(z\iv-t)m(t,z)$ with some function $m(t,z)$ which is analytic on $U_\eps$. From this
the assertion follows.
\end{proof}

Since the function $d_t$ depends analytic on $t\in \Omega$, the entries of the matrix $T_n(d_t)$ 
are also analytic in $t$. In regard to the matrix $T_n(c_t)$ we decompose for $0<t<1$
\be\label{f.10}
T_n(c_t)= T_n(e_t^+)+ K_{n,t}^+,
\ee
where $K_{n,t}^+$ is the $n\times n$ Toeplitz matrix with entries
$$
\left[K_{n,t}^{+}\right]_{jk}=
\left\{\ba{ll} 0 &\mbox{ if } j\le k\\ t^{j-k-1} &\mbox{ if } j>k.\qquad\ea\right. 
$$ 
This decomposition holds since $K_{n,t}$ is the Toeplitz matrix with the generating function equal to
$(e^{-ix}-t)\iv$ for $|t|<1$.
The right hand side of (\ref{f.10}) is well-defined and analytic for $t\in \Omega$, and hence we 
can define
$$
B_{n,t}:= T_n(e_t^+)+ K_{n,t}^+,
$$
which is the analytic continuation of $T_n(c_t)$ onto $\Omega$.
{}From this it is easy to see that the analytic continuation of $T_n(\phi)$ onto $t\in \Omega$ is 
the matrix
$$
\hat{B}_{n,t}=\twotwo{B_{n,t}}{T_n(d_t)}{T_n(d_t)^T}{B_{n,t}^T},
$$
where $A^T$ stands for the transpose of $A$.

\begin{theorem}
For each $t\in \Omega$, we have 
\bq
\lim_{n\to\iy} \det \hat{B}_{n,t} = E_t\quad\mbox{with}\quad
E_t= \frac{t}{2t(2+t^{2})+(1+2t^{2})\sqrt{2+t^{2}}}.
\eq
Moreover, the convergence is locally uniform on $\Omega$.
\end{theorem}
\begin{proof}
We make the decomposition $\phi=\phi_++\phi_0$ with
$$
\phi_+=\twotwo{e_t^+}{d_t}{\tilde{d}_t}{\tilde{e}_t^+},\qquad
\phi_0(e^{ix})=\twotwo{(e^{-ix}-t)\iv}{0}{0}{(e^{ix}-t)\iv}.
$$
In view of the definition of $B_{n,t}$ we obtain
$$
\hat{B}_{n,t}=T_n(\phi_+)+\hat{K}_n^+,\qquad \hat{K}_n^+:=
\twotwo{K_{n,t}^+}{0}{0}{(K_{n,t}^+)^T}.
$$
Now let
$$
\Theta_+(e^{ix})=\twotwo{1-te^{ix}}{0}{0}{1-te^{-ix}}.
$$
A straightforward computation shows that
$$
T_n(\Theta_+) \hat{K}_n^+ = T_n(\Theta_+\phi_0),\quad (\Theta_+\phi_0)(e^{ix})=\twotwo{e^{ix}}{0}{0}{e^{-ix}}.
$$
Moreover,
$$
T_n(\Theta_+)T_n(\phi_+)=T_n(\Theta_+\phi_+)-P_nH(\Theta_+)H(\widetilde{\phi}_+)P_n-W_n H(\widetilde{\Theta}_+)H(\phi_+)W_n.
$$
Adding the above two equations, taking the determinant and noting that 
$\det T_n(\Theta^+)=1$, it follows that 
\be\label{f.dM}
\det \hat{B}_{n,t} = \det\Big(T_n(\widehat{\phi})+P_nKP_n+W_nLW_n\Big),\qquad
\widehat{\phi}:=\Theta_+\phi_++\Theta_+ \phi_0=\Theta_+\phi.
\ee
The function $\widehat{\phi}$ belongs to $\B^{2\times 2}$ and depends analytically on $t\in\Omega$.
Moreover, $\det \widehat{\phi}=1$, hence $\wh{\phi}\in \mathcal{G}\B^{2\times 2}$ for each $t\in\Omega$. For $0<t<1$, Lemma \ref{l3.1} implies that $\wh{\phi}\in \mathcal{G}_1\B^{2\times 2}$,
and since $\Omega$ is connected this holds for each $t\in\Omega$.
The operators $K$ and $L$ are trace class operators (in fact, rank one operators), which also depend analytically on $t\in \Omega$. Hence all the assumptions of Theorem \ref{t3.1} are fulfilled,
and applying it yields the desired assertion. 
\end{proof}


\section*{Appendix A: The basic identity}

As mentioned in the introduction, the block matrix considered by the authors in \cite{FMS} is of the form
\begin{eqnarray}\label{eq2.1b}
M_n= \twotwo{\cal{R}}{\cal{Q}}{\cal{Q}}{\cal{R}}
\end{eqnarray}
where  $\cal{R}$ and$\cal{Q}$ are $n \times n$ matrices with 
entries given by 
\bq
 \cR_{jk} &=&  2(-1)^{[(k-j)/2]}R_{k-j+1} +\theta(j-k)t^{j-k-1}\nn \\
 \cQ_{jk} &=& 2i(-1)^{[(j+k)/2]}Q_{n+1-j-k}\nn
\eq
($1\le j,k\le n$), and the expressions  $R_{k}, Q_{k},$ and $\theta(k)$ are defined by the following.\\
For even $k$ 
\[R_{k} = \frac{1}{8\pi^{2}}\int_{-\pi}^{\pi}\int_{-\pi}^{\pi}\frac{\cos y \cos(kx + y)}{\cos^{2}x  + \cos^{2}y + t^{2}\cos^{2}(x + y)}dydx,\] 
and $Q_{k} = 0.$\\

\noi For odd $k$
\[
Q_{k} =  \frac{1}{8\pi^{2}}\int_{-\pi}^{\pi}\int_{-\pi}^{\pi}\frac{\cos x \cos kx}{\cos^{2}x + \cos^{2}y + t^{2}\cos^{2}(x + y )}dydx,\]
and
\[R_{k}  =  \frac{t}{8\pi^{2}}\int_{-\pi}^{\pi}\int_{-\pi}^{\pi}\frac{\cos(x+y)\cos(kx+y)}{\cos^{2}x + \cos^{2}y + t^{2}\cos^{2}(x + y )}dydx.\]
The expression $\theta(k)$ equals $1$ for $k > 0$ and $0$ otherwise.

We begin with the following lemma which describes the above coefficients as Fourier coefficients of certain functions. This is our starting place, since once this is  done, we use some simple matrix algebra to express the determinant of the above matrix as the determinant of a block Toeplitz matrix.
Throughout what follows we will assume that $0<t<1$ because
the second part of the matrix $\cal{R}$ becomes unbounded in the limit 
$n\to\iy$ in the case $t=1$.

\begin{lemma}\label{ Lemma 2.1}
Consider the functions
\[ S(e^{ix}) = \frac{t}{4\pi}\int_{-\pi}^{\pi}\frac{\cos(x + y -\pi/2) e^{i(x + y -\pi/2)}}
{\cos^{2}(x - \pi/2) + \cos^{2}y + t^{2}\cos^{2}(x + y -\pi/2)}dy ,\]
\[T(e^{ix}) = -\frac{1}{4\pi}\int_{-\pi}^{\pi}\frac{\cos y e^{i(x + y)}}
{\cos^{2}(x - \pi/2) + \cos^{2}y + t^{2}\cos^{2}(x + y -\pi/2)}dy .\]
These functions have Fourier coefficients, $S_{k}$ and $T_{k}$,
satisfying $S_{k} + T_{k} = (-1)^{[-\frac{k}{2}]}R_{-k +1}$ where $R_{k}$ is as defined as above.
\end{lemma}
\begin{proof}
Replacing $x$ by $x + \pi$ in the integral defining $S$ shows that $S(e^{i(x + \pi)}) = S(e^{ix})$ and hence that  $S_{k} = 0$ for odd $k$.  For $k$ even we have
\[ S_{k} = \frac{t}{8\pi^{2}}\int_{-\pi}^{\pi}\int_{-\pi}^{\pi}\frac{\cos(x + y -\pi/2) e^{i(x + y -\pi/2)}e^{-ikx}}
{\cos^{2}(x - \pi/2) + \cos^{2}y + t^{2}\cos^{2}(x + y -\pi/2)}dydx .\]
If we replace $x$ with $x + \pi/2$ then this integral becomes
\[\frac{t(-1)^{\frac{k}{2}}}{8\pi^{2}}\int_{-\pi}^{\pi}\int_{-\pi}^{\pi}\frac{\cos(x + y) e^{i (y - (k -1)x) }}
{\cos^{2}x + \cos^{2}y + t^{2}\cos^{2}(x + y)}dydx \]
\[ = \frac{t(-1)^{\frac{k}{2}}}{8\pi^{2}}\int_{-\pi}^{\pi}\int_{-\pi}^{\pi}\frac{\cos(x + y) \cos(y -(k-1)x)}
{\cos^{2}x  + \cos^{2}y + t^{2}\cos^{2}(x + y)}dydx .\]
The last equality follows since the inner integral as a function of $x$ is even
and this is exactly $(-1)^{-\frac{k}{2}}R_{-k+1}$ for even $k$.
If we replace $x$  by $x + \pi$ in the definition of $T$ shows that the even
coefficients vanish. A very similar computation to the above shows that for odd
$k,$ $T_{k} = (-1)^{[-\frac{k}{2}]}R_{-k+1}$ and thus combining the two cases
we have $S_{k} + T_{k} = (-1)^{[-\frac{k}{2}]}R_{-k +1}$ for all $k$.
\end{proof}

\begin{lemma}\label{ Lemma 2.2}
For $0<t<1$, the matrix ${\cal R}$ is an $n\times n$ Toeplitz matrix with symbol $ 2(S + T) + U$ where $ S$ and $T$ are defined in Lemma 1 and $U(e^{ix}) = \frac{1}{e^{-ix} - t}.$
\end{lemma}
 
\begin{proof} This follows directly from Lemma 1 and the fact that the Fourier
coefficients of $\frac{1}{e^{-ix} - t}$ are $t^{k-1}\theta(k).$
\end{proof}

\begin{lemma}\label{ Lemma 2.3}
Define 
\[V(e^{ix}) = (-1)^{[\frac{n}{2}] +1} \frac{1}{4\pi}\int_{-\pi}^{\pi}
\frac{\cos(x - \pi/2)}{\cos^{2}(x - \pi/2) + \cos^{2}y + t^{2}\cos^{2}(x + y -\pi/2)}dy.\]
Then $V_{k}$ is zero for $k$ even and for $k$ odd we have 
$V_{k} = i(-1)^{[\frac{n+1-k}{2}]}Q_{k}$
\end{lemma}
\begin{proof} To see that $V_{k}$ is zero for $k$ even, replace $x$ by $x + \pi$
as before. For the remaining equation we have
\[V_{k} = (-1)^{[\frac{n}{2}] +1} \frac{1}{8\pi^{2}}\int_{-\pi}^{\pi}\int_{-\pi}^{\pi}
\frac{\cos(x - \pi/2)e^{-ikx}}{\cos^{2}(x - \pi/2) + \cos^{2}y + t^{2}\cos^{2}(x + y -\pi/2)}dydx.\]
Making the substitution $x + \pi/2$ for $x$ and simplifying yields the result.
\end{proof}

Our next step is to take the functions given by these integal representations and simplify them.
To this end, notice that the denominator in all three integrals,
\[ \cos^{2}(x - \pi/2) + \cos^{2}y + t^{2}\cos^{2}(x + y -\pi/2)\]
can be  written as
\[\sin^{2}x + \cos^{2}y +t^{2}\sin^{2}(x+y)\]
which is in turn
\[\frac{1}{2}((2 + t^{2} -\cos 2x) +(1 - t^{2}\cos 2x) \cos 2y + t^{2}\sin 2x \sin 2y).\]
If we combine the integrals for $S$ and $T$ and use the obvious trigonometric identities
we see that the numerator is of the form 
\[\frac{1}{2}((t - e^{ix}) - (te^{2ix} + e^{ix})\cos 2y -i(te^{2ix} + e^{ix})\sin 2y )\]
 Thus combining $S$ and $T$ we see that we have an integral
 of the form
 \[\frac{1}{4\pi}\int_{0}^{2\pi}\frac{D + E\cos 2y +F\sin2y}{A + B\cos 2y +C\sin2y}dy,\]
 where the coefficients depend on $x$ and $t$ and are given by the above two
 equations for the numerator and denominator.
 
 It is fairly straightforward to evaluate such an integral. First we replace $2y$ by
 $y$ which results in no change because of periodicity. Second we define
 \[ \cos \phi = \frac{B}{\sqrt{B^{2} + C^{2}}}, \,\,\,\,\,\,\sin \phi = \frac{C}{\sqrt{B^{2} + C^{2}}}\]
 and change variables using $y = z + \phi.$ At this point we have an integral of the form
 \[ \frac{1}{4\pi}\int_{0}^{2\pi}\frac{D + G\cos z + H \sin z}{A + K\cos z} dz .\]
 The contribution from the $\sin$-term is zero and the rest can be easily evaluated
 since it reduces to one of the form 
 $$\int_{0}^{2\pi} \frac{dz}{ 1 + a \cos z} = \frac{2\pi}{\sqrt{1 - a^{2}}}$$ providing that 
 $|a| < 1.$ Using this equation we can compute the integral for $S + T$.
 In the same way we are also able to evaluate the integral for the 
 function $V$. This computation is even more straightforward
 and we leave the details to the reader. Notice 
 that the denominator in the defining integral is the same as for $S$ and $T$ and the numerator only  depends on $x$.
 
 \begin{lemma}\label{ Lemma 2.4} We have
\bq
S + T &=& -\frac{t \cos x + \sin^{2} x}{2(t - e^{-ix})\sqrt {t^{2 } + \sin^2 x+\sin^4x}} + \frac{1}{2(t - e^{-ix})},
\nn\\
V &=& \frac{ (-1)^{[n/2]+1}\sin x}{2\sqrt {t^{2 } + \sin^2 x+\sin^4x}}.\nn
\eq
\end{lemma}
Hence the generating function for the Toeplitz matrix $\cal{R}$ is given by
\[2(S+T)  + U= \frac{ t\cos x + \sin^{2}x}{(e^{-ix}-t)\sqrt{t^{2 } +\sin^2 x+\sin^4x}}.\]
Notice finally that the ($j,k$)-entry ($1\le j,k\le n$) of the $\cal{Q}$ matrix is given by twice the $n+1-j-k$ Fourier coefficient of $V$. 

Now that we have identified the entries in each of the blocks of the matrix (\ref{eq2.1b}) as Fourier coefficients of certain functions, it is fairly easy to see how the determinant of (\ref{eq2.1b}) can be computed as the determinant of a block Toeplitz matrix.
As a preliminary step, we define an operator $W_{n}$ on the finite-dimensional complex vector space of dimension $n$ by 
\[W_{n}(a_{1}, \dots , a_{n}) = (a_{n}, \dots, a_{1})\]
and denote the identity operator on the same space by $I_{n}.$ Multiplication of a matrix  by $W_{n}$ on the right results in changing the ($j,k$)-entry into the ($j,n+1-k$)-entry  and on  the left changes the entry to the ($n+1-j,k$)-entry. Also note that $\det W_{n}^{2} = 1.$
Hence the block matrix product
\[ \twotwo{I_{n}} {0}{0}{ W_{n}} \twotwo{\cal{R}}{\cal{Q}}{\cal{Q}}{\cal{R}} \twotwo{I_{n}} {0}{0}{ W_{n}}
\]
has the same determinant as the one given in formula (\ref{eq2.1b}) and thus we can replace it with the product. The advantage to this is that 
since 
\[ \twotwo{I_{n}} {0}{0}{ W_{n}} \twotwo{\cal{R}}{\cal{Q}}{\cal{Q}}{\cal{R}} \twotwo{I_{n}} {0}{0}{ W_{n}}
= \twotwo{\cal{R}}{{\cal{Q}}W_{n} }{W_{n}\cal{Q}} {W_{n} {\cal{R} }W_{n}} \]
the ($j,k$)-entry of ${\cal{Q}}W_{n}$ is now the $k-j$ Fourier coeffcient of the function $V$ and the ($j,k$)-entry of  $W_{n}\cal{Q}$ is $j-k$ and thus the upper-right and lower-left matrices are now Toeplitz in structure. The matrix $W_{n} {\cal{P} }W_{n}$ still remains Toeplitz, but with symbol $(2(S+T) +U)(e^{ix})$ replaced by $(2(S+T) +U)(e^{-ix})$.

The end result of this is that we have shown that (for $0<t<1$)
\bq\label{def.Tphi3}
\det M_n=\det T_n(\phi)\quad\mbox{ with} \quad \phi= \twotwo{c}{d}{{\tilde d}}{{\tilde c}} 
\eq
where 
\bq
c(e^{ix}) & = &  \frac{ t \cos x + \sin^{2}x}{(e^{-ix}-t)\sqrt{t^{2 }+ \sin^{2}x +\sin^{4}x}} \nn\\
d(e^{ix}) &=& \frac{\sin x}{\sqrt{t^{2 }+ \sin^{2}x +\sin^{4}x} },\nn
\eq
$\tilde{c}(e^{ix})=c(e^{-ix})$, $\tilde{d}(e^{ix})=d(e^{-ix})$ 
Notice that dropping the sign when passing from $V$ to $d$ does not influence the value of the 
determinant.

\vspace{3ex}\noindent
{\em Acknowledgement.}
The authors would like to thank Paul Fendley for introducing the topics of this paper to us and also Jim Delany who helped us with mathematica computations in our early investigations.


\end{document}